\documentclass[prl,showpacs,twocolumn]{revtex4}
\usepackage{mathrsfs}
\usepackage{amsmath}
\usepackage{amssymb}
\usepackage{revsymb}
\usepackage{graphicx}
\usepackage{mathrsfs}

\begin{document}
\title{Light diffraction by a strong standing electromagnetic wave}
\author{A. \surname{Di Piazza}}
\email{dipiazza@mpi-hd.mpg.de}
\author{K. Z. \surname{Hatsagortsyan}}
\email{k.hatsagortsyan@mpi-hd.mpg.de}
\author{C. H. \surname{Keitel}}
\email{keitel@mpi-hd.mpg.de}
\affiliation{Max-Planck-Institut f\"ur Kernphysik, Saupfercheckweg 1, D-69117 Heidelberg, Germany}

\date{\today}

\begin{abstract}
The nonlinear quantum interaction of a linearly polarized x-ray probe beam with a focused intense standing laser wave is studied theoretically. Because of the tight focusing of the standing laser pulse, diffraction effects arise for the probe beam as opposed to the corresponding plane wave scenario. A quantitative estimate for realistic experimental conditions of the ellipticity and the rotation of the main polarization plane acquired by the x-ray probe after the interaction shows that the implementation of such vacuum effects is feasible with future X-ray Free Electron Laser light.
\end{abstract}


\maketitle
Since the early work by Heisenberg and Euler \cite{Heisenberg_1936} the electromagnetic properties of vacuum
are known in principle to be modified by the presence of strong electromagnetic fields \cite{Dittrich_b_2000}. The associated scales for the electric and magnetic field amplitudes are governed by the so called critical electric $E_{cr}=m^2c^3/\hbar e=1.3\times 10^{16}\;\text{V/cm}$ and magnetic $B_{cr}=m^2c^3/\hbar e=4.4\times 10^{13}\;\text{G}$ fields with negative electron charge $-e$ and electron mass $m$. In the presence of such strong fields vacuum generally behaves as a nonlinear, birefringent and dichroic dielectric medium. In particular the vacuum polarization has been studied in the presence of static and uniform electromagnetic fields in various configurations \cite{Vac_Pol_Static}. Given the extremely large values of $E_{cr}$ and $B_{cr}$ it remains very challenging though to experimentally verify vacuum nonlinearities by means of static and uniform fields. The PVLAS (Polarizzazione del Vuoto con Laser) experiment has recently been designed to measure the extremely small ellipticity acquired by a linearly polarized probe laser after passing repeatedly through a vacuum region with applied static uniform magnetic field of strength $5.5\times 10^4\;\text{G}$ \cite{PVLAS}. We note that most recently first experimental results from the PVLAS project on the rotation of light polarization in vacuum have been reported in \cite{Zavattini_2006}. Nevertheless, those results cannot be explained as a nonlinear quantum electrodynamics (QED) effect but as a dichroic effect due to the possible conversion of a photon into a pseudoscalar particle, called axion.

Much stronger electromagnetic fields can be produced by means of focused laser pulses. Laser intensities up to $10^{21}\text{-}10^{22}\;\text{W/cm$^2$}$ have already been obtained in the so called $\lambda^3$ regime \cite{Bahk_2004}. Envisaged intensities of order $10^{24}\text{-}10^{26}\;\text{W/cm$^2$}$ corresponding to peak electric fields $10^{13}\text{-}10^{14}\;\text{V/cm}$ are likely to be reached in near future \cite{Tajima_2002}. At present, elastic light-light interaction has not been experimentally revealed via strong laser pulses [see also \cite{Lundstrom_2006}]. On the theoretical side, photon propagation was evaluated to be modified in the presence of an intense plane wave \cite{Laser_Pol,Aleksandrov_1986}. Recently, the analogous problem of the propagation of an x-ray photon along a standing wave has been considered in \cite{Heinzl_2006} by estimating the effects of the laser pulse profile along the propagation direction. With the recent development of extremely focused laser beams, a theoretical treatment becomes necessary though which fully takes into account of the three-dimensional spatial confinement of the fields.

In the present Letter we investigate how extreme spatial confinement of crossed laser fields gives rise to diffraction effects in the interaction between probe and intense laser beams. The virtues of a high-frequency probe field with good spatial coherence and a large photon number per pulse become apparent such that the envisaged X-ray Free Electron Laser (X-FEL) at Deutsches Elektronen-Synchrotron in Hamburg appears suitable. The required focusing of the laser pulses turns out to yield a significant reduction of the observable vacuum nonlinearities while experimental verification is shown to be feasible for realistic near future parameters.

Since vacuum polarization effects are larger with shorter probe wavelength, we study the interaction of an \emph{x-ray} probe beam by a standing wave generated by the superposition of two counterpropagating strong and tightly focused
optical laser beams. The advantage of a standing wave instead of a single laser wave is a larger coupling strength in the configuration in which the probe wave propagates perpendicularly to the strong beam. Our configuration also
turns out to be more favorable because the deteriorating role of diffraction on vacuum effects is reduced.

In the present experimental conditions it is appropriate to assume that the amplitude and the frequency of both the probe and the strong field are much less than $E_{cr}$ and $m$ respectively (from now on natural units with $\hbar=c=1$ are used). Further, we have ensured that here the axion effect is completely negligible, essentially due to the microscopic dimensions of the interaction region of the two beams [see also \cite{Raffelt_1988}]. For these reasons our starting point is the Euler-Heisenberg Lagrangian density at lowest order \cite{Dittrich_b_2000}:
\begin{equation}
\label{L}
\mathscr{L}=\frac{1}{2}(E^2-B^2)+\frac{2\alpha^2}{45m^4}\left[(E^2-B^2)^2+7(\mathbf{E}\cdot\mathbf{B})^2\right]
\end{equation}
with fine-structure constant $\alpha=e^2/4\pi$ and total electric and magnetic field $\mathbf{E}(\mathbf{r},t)$ and $\mathbf{B}(\mathbf{r},t)$, respectively. The second term in the previous Lagrangian density can be considered as a small perturbation of the Maxwell Lagrangian density $(E^2-B^2)/2$. The Lagrangian density in Eq. (\ref{L}) yields the following nonlinear wave equation for the electric field
\begin{equation}
\label{Wave_Eq}
\nabla^2\mathbf{E}-\partial_t^2\mathbf{E}=\boldsymbol{\mathcal{J}}(\mathbf{r},t)
\end{equation}
where
\begin{equation}
\label{J}
\boldsymbol{\mathcal{J}}(\mathbf{r},t)=\boldsymbol{\nabla}\times\left(\partial_t \mathbf{M}\right)+\partial_t^2\mathbf{P}-\boldsymbol{\nabla}(\boldsymbol{\nabla}\cdot\mathbf{P})
\end{equation}
with polarization  $\mathbf{P}(\mathbf{r},t)=4\alpha^2/(45m^4)[2(E^2-B^2)\mathbf{E}+7(\mathbf{E}\cdot\mathbf{B})\mathbf{B}]$ and magnetization $\mathbf{M}(\mathbf{r},t)=4\alpha^2/(45m^4)[2(E^2-B^2)\mathbf{B}-7(\mathbf{E}\cdot\mathbf{B})\mathbf{E}]$. Since the quantity $\boldsymbol{\mathcal{J}}(\mathbf{r},t)$ is very small, this nonlinearity of the wave equation (\ref{Wave_Eq}) can be accounted for by a perturbative approach. Up to first order, the solution of Eq. (\ref{Wave_Eq}) can be expressed as:  $\mathbf{E}(\mathbf{r},t)\approx \mathbf{E}^{(0)}(\mathbf{r},t)+\mathbf{E}_{\text{diff}}(\mathbf{r},t)$, where $\mathbf{E}^{(0)}(\mathbf{r},t)=\mathbf{E}_0(\mathbf{r},t)+\mathbf{E}_p(\mathbf{r},t)$ is the zero-order solution with $\mathbf{E}_0(\mathbf{r},t)$ being the strong standing wave electric field and $\mathbf{E}_p(\mathbf{r},t)$ the probe electric field and where
\begin{equation}
\label{E_1}
\mathbf{E}_{\text{diff}}(\mathbf{r},t)=-\frac{1}{4\pi}\int_V d\mathbf{r}'\frac{\boldsymbol{\mathcal{J}}^{(1)}(\mathbf{r}',t-|\mathbf{r}-\mathbf{r}'|)}{|\mathbf{r}-\mathbf{r}'|}
\end{equation}
is the diffracted wave generated by the nonlinear QED interaction of the probe with the strong standing wave. This interaction breaks the space isotropy [see Eq. (\ref{L})] and the symmetry breaking manifests itself in the field correction $\mathbf{E}_{\text{diff}}(\mathbf{r},t)$. In Eq. (\ref{E_1}) it is understood that the point $\mathbf{r}$ lies outside the interaction volume $V$ defined by the condition $\boldsymbol{\mathcal{J}}^{(1)}(\mathbf{r},t)\neq\mathbf{0}$. $\boldsymbol{\mathcal{J}}^{(1)}(\mathbf{r},t)$ is obtained by substituting the fields in Eq. (\ref{J}) by their corresponding zero-order expressions $\mathbf{E}^{(0)}(\mathbf{r},t)$ and $\mathbf{B}^{(0)}(\mathbf{r},t)$. Further, we assume that the strong standing wave results from the superposition of two Gaussian beams propagating along the $z$ axis in opposite directions and both with polarization along the $x$ axis, amplitude $E_0/\sqrt{2}$, frequency $\omega_0$ and waist size $\text{w}_0$ (see Fig. 1) \cite{Salamin_2002} yielding
\begin{figure}
\begin{center}
\includegraphics[width=6cm]{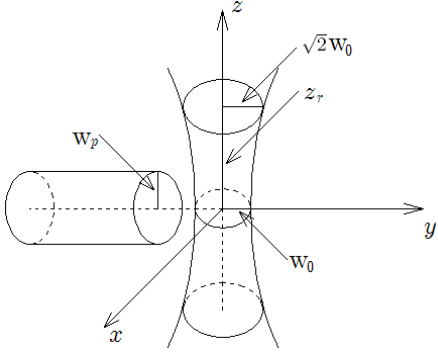}
\end{center}
\caption{Geometry depicting the interaction of the two beams. All symbols are described between Eqs. (\ref{E_1}) and (\ref{E_1_omega}).}
\end{figure}
\begin{equation}
\label{E_0}
\begin{split}
\mathbf{E}_0(\mathbf{r},t)&=\sqrt{2}E_0\frac{e^{-(x^2+y^2)/\text{w}^2(z)}}{\sqrt{1+(z/z_r)^2}}\sin(\psi_0+\omega_0t)\\
&\quad\times\cos\left(k_0z-\tan^{-1}\frac{z}{z_r}+\frac{k_0z}{2}\frac{x^2+y^2}{z^2+z_r^2}\right)\hat{\mathbf{x}}
\end{split}
\end{equation}
with $k_0=\omega_0$, $\text{w}(z)=\text{w}_0\sqrt{1+(z/z_r)^2}$ and Rayleigh length $z_r=k_0\text{w}_0^2/2$. The factor $1/\sqrt{2}$ in the amplitude of the superimposed beams has been inserted to take into account that usually in experiments the standing wave is obtained by first splitting the beam of one laser field. In general, Eq. (\ref{E_0}) is nearly an approximate solution of Maxwell's equations when $\text{w}_0/z_r\ll 1$. We have shown that our final results are valid up to terms proportional to $(\text{w}_0/z_r)^2$ which is much smaller than one even in the case of a maximally focused beam with $\text{w}_0=\lambda_0=2\pi/k_0$. As opposed to the strong standing wave, the probe beam usually is neither tightly focused nor very strong. For these reasons, if we assume that the probe field propagates along the $y$ axis and that it is polarized in the $x$-$z$ plane, we can write the probe electric field as
\begin{equation}
\label{E_p}
\begin{split}
\mathbf{E}_p(\mathbf{r},t)&=E_pe^{-(x^2+z^2)/w_p^2}\sin\left(\psi_p+\omega_pt-k_py\right)\\
&\quad\times(\hat{\mathbf{z}}\sin\vartheta+\hat{\mathbf{x}}\cos\vartheta)
\end{split}
\end{equation}
with probe frequency $\omega_p=k_p$ and probe waist size $\text{w}_p$ (see Fig. 1). In Fourier space Eq. (\ref{E_1}) becomes:
\begin{equation}
\label{E_1_omega}
\mathbf{E}_{\text{diff}}(\mathbf{r},\omega)=-\frac{1}{4\pi}\int_V d\mathbf{r}'\frac{e^{i\omega|\mathbf{r}-\mathbf{r}'|}}{|\mathbf{r}-\mathbf{r}'|}\boldsymbol{\mathcal{J}}^{(1)}(\mathbf{r}',\omega).
\end{equation}
We are interested in the effects of the strong standing wave on the probe field. Then we fix $\omega=\omega_p$ and the detection point $\mathbf{r}$ along the probe field propagation axis, i. e. $\mathbf{r}=\mathbf{r}_d\equiv(0,y_d,0)$ with $y_d>\text{w}_0$ (see Fig. 1). Moreover, to analyze the diffracted field evolution along its propagation to $\mathbf{r}_d$, we write a general expression for the diffracted field which is valid in the near as well as in the far zone as defined below. This can be realized by adopting the following conditions: on the one hand $y_d\gg \text{w}_p,\text{w}_0,z_r$ and on the other $(\text{w}_0/\lambda_p)(\text{w}_p/y_d)^2\ll 1$ and $(\text{w}_p/\lambda_p)(\text{w}_p/y_d)^3\ll 1$, with probe wavelength $\lambda_p=2\pi/\omega_p$. From an experimental point of view, the previous conditions are not very restrictive. For example, if $\text{w}_0\sim\lambda_0\lesssim 1\;\text{$\mu$m}$, $y_d\gtrsim 1\;\text{cm}$ and $\lambda_p\gtrsim 10^{-3}\;\text{$\mu$m}$, these inequalities are fulfilled if $\text{w}_p\lesssim 100\;\text{$\mu$m}$. Now, if we set
\begin{equation}
\label{E_diff_t}
\mathbf{E}_{\text{diff}}(\mathbf{r}_d,t)=-\mathbf{E}_{\text{diff}}(\mathbf{r}_d)\frac{e^{-i(\psi_p+\omega_pt-k_py_d)}}{2i}+\text{c.c.},
\end{equation}
we obtain for above parameter region
\begin{equation}
\label{E_diff}
\mathbf{E}_{\text{diff}}(\mathbf{r}_d)=\frac{\omega_p^2}{4\pi}
\frac{E_p}{2}(7\sin\vartheta\hat{\mathbf{z}}+4\cos\vartheta\hat{\mathbf{x}})
\frac{\alpha}{45\pi}\left(\frac{E_0}{E_{cr}}\right)^2\frac{\mathscr{V}}{y_d}
\end{equation}
where the complex integration
\begin{equation}
\label{V}
\begin{split}
\mathscr{V}&= \int_V d\mathbf{r}' e^{-(1/\text{w}_p^2-i\omega_p/2y_d)(x^{\prime 2}+z^{\prime 2})}\frac{e^{-2(x^{\prime 2}+y^{\prime 2})/\text{w}^2(z')}}{1+(z'/z_r)^2}\\ 
&\;\;\;\times\left[1+\cos 2\left(k_0z'-\tan^{-1}\frac{z'}{z_r}+\frac{k_0z'}{2}\frac{x^{\prime 2}+y^{\prime 2}}{z^{\prime 2}+z_r^2}\right)\right]
\end{split}
\end{equation}
should be performed over the finite volume $V$ of the interaction region. Furthermore, the three-dimensional integral over $\mathbf{r}'$ is rapidly convergent in the limit $V\to \infty$ so that we can evaluate it in this limit with no appreciable error. Hence, with $\text{w}_p/\lambda_0\gg \sqrt[4]{1+(\pi \text{w}_p^2/y_d\lambda_p)^2}$, the $\cos$-term in the integrand yields a negligible contribution with respect to the remaining integral that can be performed exactly. For example, if $\lambda_0\lesssim 1\;\text{$\mu$m}$, $\text{w}_p\gtrsim 8\;\text{$\mu$m}$ and $\lambda_p\gtrsim 10^{-3}\;\text{$\mu$m}$ this approximation applies well in the region $y_d\gtrsim 2\;\text{cm}$. Then we obtain
\begin{equation}
\label{V_a}
\begin{split}
\mathscr{V}= & \frac{\pi}{\sqrt{2}}\frac{\text{w}_0 \text{w}_p z_r}{\sqrt{1-i\pi\frac{\text{w}_p^2}{y_d\lambda_p}}}\exp\left[z_r^2\left(\frac{1}{2}\frac{1}{\text{w}_p^2}+\frac{1}{\text{w}_0^2}-i\frac{\pi}{2}\frac{1}{y_d\lambda_p}\right)\right]\\
\qquad&\times\text{K}_0\left[z_r^2\left(\frac{1}{2}\frac{1}{\text{w}_p^2}+\frac{1}{\text{w}_0^2}-i\frac{\pi}{2}\frac{1}{y_d\lambda_p}\right)\right]
\end{split}
\end{equation}
with $\text{K}_0(z)$ being the zero-order modified Bessel function \cite{Ryzhik_b_1965}. By adding the diffracted field to the probe field in Eq. (\ref{E_p}) we obtain that the resulting field is elliptically polarized in the $x\text{-}z$ plane and that the major axis of the ellipse is rotated with respect to the initial probe polarization direction. The two relevant parameters, i. e. the rotation angle of the major axis of the ellipse and the ellipticity, are given respectively by
\begin{align}
\label{psi}
\psi &=\frac{\omega_p}{8\pi}\frac{3\alpha}{45\pi}\frac{I_0}{I_{cr}}\frac{\omega_p\mathscr{V}_r}{2y_d}\sin 2\vartheta,\\
\label{epsilon}
\varepsilon &=\frac{\omega_p}{8\pi}\frac{3\alpha}{45\pi}\frac{I_0}{I_{cr}}\frac{\omega_p\mathscr{V}_i}{2y_d}\sin 2\vartheta
\end{align}
with the real (imaginary) part $\mathscr{V}_r$ ($\mathscr{V}_i$) of $\mathscr{V}$, strong laser intensity $I_0$ and $I_{cr}=E_{cr}^2/8\pi=2.3\times 10^{29}\;\text{W/cm$^2$}$. Since the diffracted field $\mathbf{E}_{\text{diff}}(\mathbf{r},t)$ is generated by a localized source inside $V$, it is not a plane wave and its amplitude depends on the observation distance $y_d$ as $\psi$ and $\varepsilon$ do.

Eqs. (\ref{psi}) and (\ref{epsilon}) and the analytical expression (\ref{V_a}) allow to analyze the evolution along the propagation direction of the polarization of the probe field after the diffraction by the strong standing wave. The typical length of the interaction region is $\text{w}_0$ in the $x$ direction and $\text{w}_p$ in the $z$ direction and they determine the diffraction parameters: $\xi_0=\text{w}_0^2/y_d\lambda_p$ and $\xi_p=\text{w}_p^2/y_d\lambda_p$. In turn, these determine the field zones: the ``near zone'', if $\xi_0,\xi_p\gg 1$, where the diffraction effects along both the $x$ and $z$ axis are negligible; the ``far zone'', if $\xi_0,\xi_p\ll 1$, where the diffraction effects are very important. 

Only in the near zone one obtains results analogous to those in \cite{Laser_Pol,Heinzl_2006} because the spatial confinement of the fields transverse to the probe propagation direction does not play any role. In this zone, on one hand the rotation angle $\psi$ is much smaller than the ellipticity being the dominating imaginary part of $\mathscr{V}\approx i\sqrt{2\pi^3}\text{w}_0y_d/\omega_p$ in Eq. (\ref{V_a}). On the other hand, the ellipticity does not depend on the observation distance and it can be written as $2\varepsilon=\omega_pl(n_{\perp}-n_{\parallel})\sin 2\vartheta$. In this expression $l$ is the distance covered by the probe in the presence of the strong field and $n_{\parallel,\perp}$ are two different vacuum refractive indices depending on the mutual orientation of the probe and the strong field polarization directions. In our case $l=2\text{w}_0$ and $n_{\parallel,\perp}=1+\sqrt{\pi/2}c_{\parallel,\perp}\alpha/(360\pi)I_0/I_{cr}$ with $c_{\parallel}=4$ and $c_{\perp}=7$. However, we stress that the near zone is hardly realizable experimentally. For example, if $\text{w}_0\lesssim 1\;\text{$\mu$m}$, the condition $\xi_0\gg 1$ requires observation distances $y_d\ll 1\;\text{cm}$ even in the case of x-ray probes.

In the far zone where $\xi_0,\xi_p\ll 1$, $\mathscr{V}$ is independent of $y_d$ and becomes real. Then the polarization of the probe field remains linear but its polarization direction is rotated appreciably by $\psi$. In particular, if $\text{w}_p\gg \text{w}_0$ then $\mathscr{V}\approx \pi^{3/2}\text{w}_p\text{w}_0^2/2$. Therefore, the polarization rotation angle $\psi$ in the far zone is $\pi\sqrt{\xi_p\xi_0/2}$ times smaller than the ellipticity in the near zone. However, we note that at observation distances with $\xi_p<1/\pi$ the defocusing of the probe field is no more negligible.  For a precise rather than above order-of-magnitude estimate correcting terms proportional to $1/(\pi\xi_p)$ should be included in Eq. (\ref{E_p}).

In the remaining intermediate zone the most general situation happens in which the polarization rotation angle $\psi$ and the ellipticity $\varepsilon$ alter similarly. In Fig. 2 we plot $\psi$ and $\varepsilon$ as functions of the observation distance $y_d$ and find detectable values. In the numerical estimates we have used the exact expression (\ref{V}) of $\mathscr{V}$ and ensured that the analytic expression (\ref{V_a}) reproduces the numerical values up to $0.1\;\%$. 
\begin{figure}
\begin{center}
\includegraphics[width=7cm]{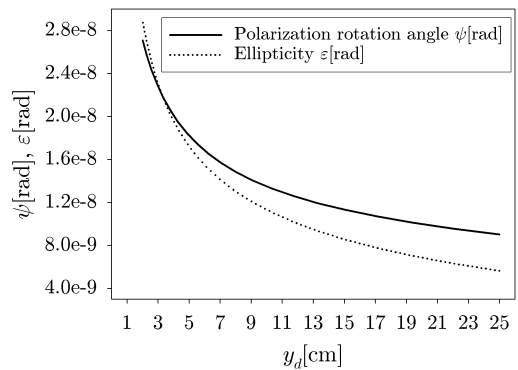}
\end{center}
\caption{The polarization rotation angle $\psi$ and the ellipticity 
$\varepsilon$ as functions of the observation distance $y_d$. 
We set $\vartheta=\pi/4$, $w_0=\lambda_0=0.745\;\text{$\mu$m}$, 
$I_0=10^{23}\;\text{W/cm$^2$}$, $w_p=8\;\text{$\mu$m}$ and 
$\lambda_p=0.4\;\text{nm}$, yielding $\xi_0=0.14/(y_d[\text{cm}])$ 
and $\xi_p=16/(y_d[\text{cm}])$.}
\end{figure}
In order to maximize the vacuum effects, we set $\vartheta=\pi/4$. Concerning the strong field we employ the feasible intensity $I_0=10^{23}\;\text{W/cm$^2$}$ and $\text{w}_0=\lambda_0=0.745\;\text{$\mu$m}$. With respect to the probe field we set $\lambda_p=0.4\;\text{nm}$ and $\text{w}_p=8\;\text{$\mu$m}$. Note that in Fig. 2 we avoided both the far zone as mentioned above and the near zone because already at the quite small observation distance $y_d=2\;\text{cm}$ we merely achieve $\xi_0=0.07$. Nevertheless, Fig. 2 confirms that at small $y_d$ the ellipticity is larger than the polarization rotation angle and vice versa for large $y_d$. In general, as expected, the vacuum effects are larger at small $y_d$ when the effects of the diffraction along $z$ become smaller. This reduction is intuitively clear because diffraction induces a spreading and consequently an attenuation of the generated field. Furthermore, diffraction effects along $x$ can hardly be avoided. In this respect, the head-on collision of a probe field with a single laser pulse is less favorable than our crossed beam scenario because the diffraction effects then cannot be neglected along both axes $x$ and $z$.

In the plane wave approximation one neglects both the diffraction effects and the spatial confinement of the probe and the strong beams. In this case we note that the value of the ellipticity acquired by a probe with wavelength $\lambda_p=0.4\;\text{nm}$ after crossing a region with $2\text{w}_0=1.5\;\text{$\mu$m}$ and standing wave intensity $I_0=10^{23}\;\text{W/cm$^2$}$ is $4\times 10^{-7}\;\text{rad}$, i. e. more than one order of magnitude larger than our results including spatial confinement and diffraction. Nevertheless, Fig. 2 also shows that with above parameters and despite diffraction, the polarization rotation angle and the ellipticity are still more than two orders of magnitude larger than the estimated ellipticity induced by nonlinear QED effects in the PVLAS setup with $\varepsilon_{\text{PVLAS}}=5\times 10^{-11}\;\text{rad}$ \cite{PVLAS}. We stress also that our method is single-passage which does not require an optical cavity as in the PVLAS experiment.

The presence of charged particles in the interaction region may conceal vacuum effects. The maximum pressure $P_M$ of an electron gas in the interaction region to render the effects of Thompson scattering of the probe field negligible can be estimated as  $P_M\sim 10^{-6}\;\text{torr}$ for the parameters in Fig. 2 and temperature $T=300\;\text{K}$. Such a high-quality vacuum are obtained nowadays [see, e. g., \cite{Zavattini_2006}].

Another question is if todays x-ray polarimeters can measure small ellipticities and/or polarization rotation angles as obtained above. By exploiting multiple Bragg reflections by channel-cut crystals, polarimeter sensitivities of order $10^{-6}\text{-}10^{-5}\;\text{rad}$ can be reached in principle \cite{Alp_2000}. Since $\psi,\varepsilon\propto I_0$ [see Eq. (\ref{epsilon})], to obtain such values, strong field intensities of order $10^{25}\text{-}10^{26}\;\text{W/cm$^2$}$ are required which are quite feasible in the near future \cite{Tajima_2002}. Furthermore, at photon wavelengths $\lambda_p=0.4\;\text{nm}$ typical values of the modulation factor and of the efficiency of an x-ray detector are around $20\text{-}30\;\%$ \cite{Soffitta}. Then, in order to measure, for example, polarization rotation angles $\psi\sim 10^{-5}\;\text{rad}$ the probe beam should provide for each pulse at least of order $10^{12}\text{-}10^{13}$ photons. Moreover, due to the small angular acceptance of the channel-cut crystals $\Delta\theta\sim 1\;\text{$\mu$rad}$, a highly collimated x-ray probe with beam divergence up to $\Delta\theta$ is required in order to reach the mentioned values of sensitivities. Then, the probing of vacuum nonlinearities can represent a further challenging application for the future X-FEL because it is likely to fulfill both previous conditions on the number of photons per pulse and on the beam divergence \cite{Tesla}. Finally, the initial polarization degree of the X-FEL has to be sufficiently high to allow for the measurement of polarization rotation angles $\psi$ of order of $10^{-5}\;\text{rad}$. If this is not the case, the X-FEL beam needs to be further polarized by employing, for example, the technique described in \cite{Alp_2000}.

In conclusion, we have shown that diffraction and spatial confinement of the fields are essential in describing the interaction between a strong tightly focused optical standing wave and an x-ray probe beam. The ellipticity and the polarization rotation angle acquired by the probe after the interaction with the strong field were evaluated analytically. We have demonstrated that while these variables are considerably smaller than those estimated in the usually considered plane wave approximation, they should be measurable in the near future.
%
%

\end{document}